\documentclass{moriond}
\usepackage{lineno}  

\def\be{\begin{equation}}
\def\ee{\end{equation}}
\def\bea{\begin{eqnarray}}
\def\eea{\end{eqnarray}}

\newcommand{\Photo}{}

\begin{document}
\vspace*{4cm}
\title{Tau and dark sector searches at Belle and Belle II}

\author{ Wenzhe Li \\ on behalf of the Belle and Belle II collaborations}

\address{Beihang University, Beijing 102206, Peoples's Republic of China}

\maketitle

\abstracts{The Belle and Belle II experiments provide a powerful platform for beyond-the-Standard-Model searches, particularly in tau and dark sectors. This summary reports include a measurement of CP violation in $\tau^- \to \pi^- K_S^0 \nu_{\tau}$, a search for $\tau^- \to \ell^- \eta$, a search for axion-like particles via $e^+ e^- \to \gamma a (\to \gamma\gamma)$ using Belle II data, and a search for feebly interacting particles in $B \to h X_{\rm inv}$ decays using Belle data.}

\section{Introduction}
The Belle experiment~\cite{Belle:2000cnh} at KEKB~\cite{KUROKAWA20031} (1998–2010) collected 980 fb$^{-1}$, mostly at the $\Upsilon(4S)$ resonance. Its successor, Belle II~\cite{Abe:2010gxa} at SuperKEKB~\cite{Akai:2018mbz} (since 2019), features higher beam intensities, nano-beam focusing, improved triggering, tracking, and vertex resolution. To date, Belle II has recorded approximately  $0.7$ ab$^{-1}$.

The B-factories owe their name to the fact that they are optimized for the production of B-meson pairs. However the cross section for $e^+e^- \to \tau^+\tau^-$ is close to the cross section of $e^+e^- \to b\bar{b}$, so B-factories can also be considered tau factories.
This unique environment allows for precision tests of the Standard Model (SM) and searches for new physics. At the $B$-factories, $\tau$ leptons are produced in pairs in a back-to-back topology in the center-of-mass system and decay into one or three charged tracks. 
To suppress background, a tagging method is used: one $\tau$ lepton is tagged, and the signal $\tau$ is reconstructed in the opposite hemisphere, with the two hemispheres separated by a plane perpendicular to the thrust axis $\hat{t}$: $T=\max _{\hat{t}} \frac{\sum_i\left|\vec{p}_i\cdot \hat{t}\right|}{\sum_i\left|\vec{p}_i\right|}$.

The existence of dark matter is well established by astrophysical observations; however, the SM does not provide  any description of dark matter particles. Therefore, the search for the dark sector is a key physics goal of the Belle(II) experiment. The unique B-factories environment allows for highly sensitive searches for dark sector particles with masses in the MeV-GeV range.

\section{Search for CP violation in the $\tau^- \to \pi^- K_S^0\nu_{\tau}$ decays at Belle II}
In the SM, CP violation arises from a single complex phase in the Cabibbo-Kobayashi-Maskawa matrix. While well established in the quark sector, the observed magnitude falls far short of explaining the matter–antimatter asymmetry of the universe, motivating searches for additional CP violation effects, particularly in the lepton sector. 
The decay $\tau^- \to \pi^- K_S^0\nu_{\tau}$ can exhibit a nonzero CP violation decay-rate asymmetry without new physics, originating from $K^0-\bar{K}^0$ mixing interference. The SM prediction is firm $A_{C P}^{\mathrm{SM}} \simeq(0.33 \pm 0.01) \%$~\cite{Bigi:2005ts}. 
Any significant deviation would signal physics beyond the SM (BSM), such as non-standard scalar interactions.
Previous searches by BaBar and Belle showed tensions: BaBar reported a $2.8\sigma$ deviation from the SM~\cite{BaBar:2011pij}.

We report a search for CP violation in the decay $\tau^- \to \pi^- K_S^0(\geq0\pi^0)\nu_{\tau}$ using Belle II data corresponding to an integrated luminosity of 362 fb$^{-1}$. The CP violation decay-rate asymmetry is defined as:
\begin{equation}
A_{C P}=\frac{\Gamma\left(\tau^{+} \rightarrow \pi^{+} K_S^0 \nu_\tau\right)-\Gamma\left(\tau^{-} \rightarrow \pi^{-} K_S^0 \nu_\tau\right)}{\Gamma\left(\tau^{+} \rightarrow \pi^{+} K_S^0 \nu_\tau\right)+\Gamma\left(\tau^{-} \rightarrow \pi^{-} K_S^0 \nu_\tau\right)} .
\end{equation}

A $3\times1$ prong topology is employed, with one $\tau$ decaying to $\pi^- K_S^0[\to \pi^+ \pi^-](\geq0\pi^0)\nu_{\tau}$ and the other decaying leptonically ($e$ or $\mu$). A two-stage boosted decision tree (BDT) achieves a signal purity of approximately 98\%. The raw asymmetry is corrected for production/detection asymmetries, trigger effects, $K^0-\bar{K}^0$ regeneration in the detector, and residual background.

The measured asymmetry, after all corrections, is combined across $e$- and $\mu$-tagged samples: $A_1=\left(0.64 \pm 0.26_{\rm stat} \pm 0.06_{\rm syst}\right) \% $. Because the $K_S^0$ reconstruction efficiency depends on decay time, a proper comparison with the SM requires unfolding this effect~\cite{Grossman:2011zk}. The decay-time-efficiency corrected asymmetry is: $A_{C P}=\left(0.71 \pm 0.26_{\rm stat} \pm 0.06_{\rm syst} \pm 0.15_{\rm unf}\right) \% $ , as shown in Fig.~\ref{Tau}.

This result is consistent with SM prediction of $A_{CP}^{SM}$ within $1.24\sigma$. It does not confirm the upward fluctuation previously reported by the BaBar collaboration ($2.8\sigma$ deviation) and reduces the overall tension with the SM. With a precision already comparable to that of BaBar, this measurement demonstrates the strong potential of Belle II to provide a decisive test of CP violation in the $\tau$ sector with larger data samples.

\section{Search for lepton flavor violation $\tau^- \to \ell^-\eta$ decays at Belle II}

Lepton flavor violation (LFV) is heavily suppressed in the Standard Model (SM), with predicted branching fractions below $10^{-50}$ even after including neutrino oscillations. However, many BSM scenarios, such as leptoquarks, little Higgs models with $T$-parity, and type-I/III seesaw mechanisms, predict LFV $\tau$ decays at the level of $10^{-8}$, within reach of current high-luminosity $B$ factories. Previous searches by CLEO~\cite{CLEO:1997gxa}, Babar~\cite{BaBar:2006jhm}, and Belle~\cite{Belle:2007cio} found no signal, with Belle setting the most stringent limits: $9.2\times10^{-8}$ ($6.5\times10^{-8}$) for the electron (moun) channel at 90\% confidence level (C.L.). 

We present a search for $\tau^{-} \to e^{-} \eta$ and $\tau^{-} \to \mu^{-} \eta$ using \(427.9~\mathrm{fb}^{-1}\) of data collected by the Belle II detector at the SuperKEKB \(e^+e^-\) collider. The \(\eta\) meson is reconstructed in two decay modes: $\eta \to \gamma\gamma$ and $\eta \to \pi^+\pi^-\pi^0$. Events are divided into two hemispheres by the thrust axis. The
tag side is required to be a one\-prong decay. 

Background suppression uses kinematic selections and, for the $\eta\to\gamma\gamma$ modes, a gradient-BDT to reject dominant backgrounds from Bhabha, dimuon, and radiative processes. For $\eta\to\pi^+\pi^-\pi^0$, a cut-based selection is sufficient due to lower background levels. Signal candidates are identified in the two-dimensional plane of the reconstructed $\tau$ mass $M$ and energy difference $\Delta E$. Background yields are estimated using data-driven methods: an ABCD method for the $\eta\to\gamma\gamma$ channel and a two-dimensional unbinned maximum-likelihood fit for the $\eta\to\pi^+\pi^-\pi^0$ channel. 

\begin{figure}
    \centering
    \includegraphics[width=0.41\linewidth]{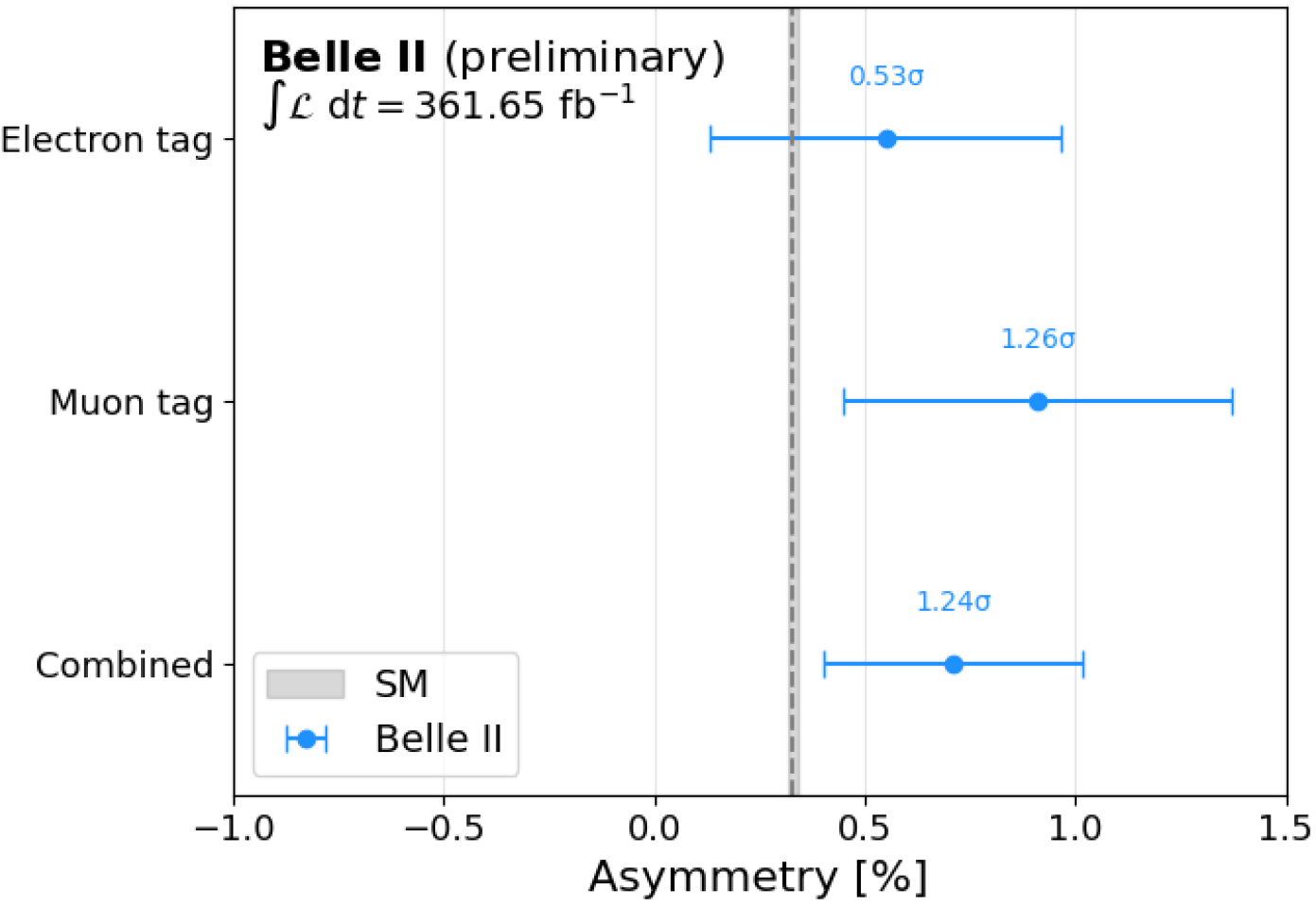}
    \includegraphics[width=0.42\linewidth]{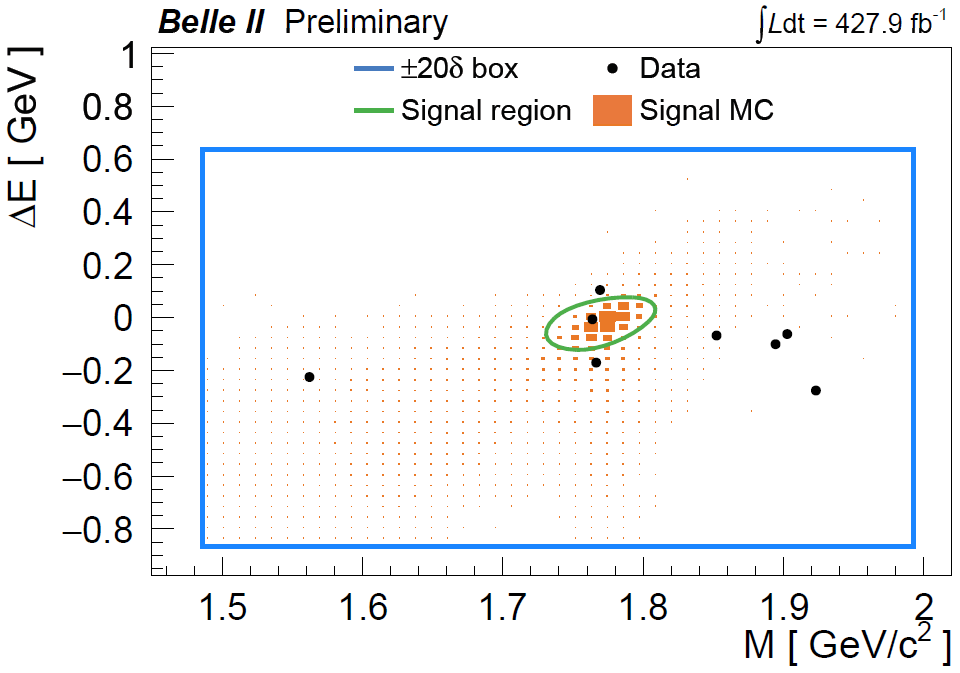}
    \caption{(Left) Efficiency-corrected CP asymmetry for the same decays for $e$-, $\mu$-tagged samples and the combined result at Belle II. (Right) Distributions of $\tau^- \to \ell^- \eta$ in the $(M, \Delta E)$ plane for $\tau^- \to e^-\eta(\to \gamma\gamma)$ mode. }
    \label{Tau}
\end{figure}

The observed event yields in the signal regions are consistent with background expectations as shown in the right plot of the Fig.~\ref{Tau}. No significant excess is observed. The upper limits on the branching fractions are set using the $\rm CL_{s}$ method, implemented in the RooStats framework. The resulting 90\% C.L. upper limits on the branching fractions are $\mathcal{B}(\tau^{-} \to e^{-} \eta) < 9.21 \times 10^{-8}$ and $\mathcal{B}(\tau^{-} \to \mu^{-} \eta) < 4.23 \times 10^{-8}$. The limit for the electron mode is consistent with the previous best result from Belle, while the limit for the muon mode is the most stringent to date. 

\section{Search for axion-like particles in $e^+e^- \to \gamma a(\to\gamma\gamma)$ at Belle II}

Axion-like particles (ALPs) appear in many extensions of the SM. Unlike the QCD axion, their mass and couplings are independent, making them a generic portal to dark sectors. This analysis focuses on ALPs that couple predominantly to photons with coupling strength $g_{a\gamma\gamma}$, assuming $\mathcal{B}(a\to\gamma\gamma)=1$. The search is performed via the ALP-strahlung process $e^+e^- \to \gamma a$ using $408~\mathrm{fb}^{-1}$ of data collected by Belle II detector.

The three most energetic photons in each event are used to form three diphoton candidates per event. A kinematic fit constraining the three-photon four-momentum to the initial state improves the diphoton mass resolution. Backgrounds, dominated by $e^+e^- \to \gamma\gamma(\gamma)$, are suppressed using a  neural network with 19 kinematic and topological variables. The selection efficiency ranges from $25\%$--$35\%$ for masses below $7~\rm{GeV}/c^2$, decreasing to $\sim10\%$ at the highest masses.

The search is performed with the signal region in the diphoton invariant mass $M_{\gamma\gamma}$ kept hidden until the analysis procedure is finalized using simulated data. For each ALP mass hypothesis $m_a$, a narrow peak is sought over the range $0.17 < m_a < 9.80~\rm{GeV}/c^2$, with step sizes approximately half the signal mass resolution. Additionally, the regions corresponding to the $\eta$ and $\omega$ regions are excluded from the analysis. The signal yield is extracted via an extended unbinned maximum-likelihood fit to the $M_{\gamma\gamma}$ distribution. The fit is performed independently for each mass hypothesis.

No significant signal excess is observed. The largest local (global) significance is $3.3\sigma$ ($1.4\sigma$) at $m_a = 0.22~\mathrm{GeV}/c^2$. Upper limits at $95\%$ C.L. are set on the cross section $\sigma(e^+e^- \to \gamma a(\to\gamma\gamma))$ and on the ALP-photon coupling $g_{a\gamma\gamma}$. 
The limits on $g_{a\gamma\gamma}$ are the most restrictive to date over nearly the entire mass range $0.17 < m_a < 5.00~\mathrm{GeV}/c^2$, as shown in the left plot of the Fig.~\ref{Dark}, improving upon previous results~\cite{Belle-II:2020jti} by up to a factor of nine. This analysis demonstrates the unique sensitivity of Belle II to light ALPs and sets a new benchmark for photon-coupled dark sector searches at $e^+e^-$ colliders.

\section{Search for feebly-interacting particles in $B$ decays with missing energy at Belle}

Feebly-interacting particles (FIPs) are predicted in many extensions of the SM. Prominent candidates include pseudoscalar ALPs, dark scalars, light vector particles (dark photons), and R-parity violating supersymmetry (RPV SUSY). In $B$ decays, FIPs can be produced via flavor-changing neutral current or charged current processes, manifesting as a missing energy signature when the particle is long-lived or decays invisibly. Such searches are complementary to direct production at $e^+e^-$ colliders and provide unique sensitivity to light dark sector mediators.

The analysis is performed using $711~\rm{fb}^{-1}$ of data collected by Belle detector at the $\Upsilon(4S)$ resonance. The search targets the decays  
$
B^+ \to h X_{\rm{inv}},\ h = \pi^+, K^+, p, D_s^+,\ B^0 \to \bar{D}^0 X_{\rm{inv}},
$
where $X_{\rm{inv}}$ is an invisible FIP. The signal-side $B$ meson is reconstructed via the hadronic system $h$, and the tag-side $B$ is fully reconstructed using the full event interpretation algorithm. The observable is the $h$ momentum in the signal $B$ rest frame, $p_h$, which exhibits a narrow peak for a two-body decay. A scan over $m_{X_{\mathrm{inv}}}$ from $\sim 1\ \mathrm{MeV}/c^2$ up to $m_B-m_h$ is performed, with a scan step equal to half the signal resolution. Mass regions for known SM particles $\pi^0$, $K^{(*)0}$, $D^{(*)0}$, $\eta_c$, $\chi_{c1}$ and $\psi(2S)$ are vetoed.

Background suppression uses a two-stage BDT: BDT1 rejects $e^+e^-\to q\bar{q}$ continuum, and BDT2 suppresses remaining $B\bar{B}$ backgrounds. The combined signal efficiency is $\sim 84\%$ with $\sim 86\%$ background rejection. The signal yield is extracted via unbinned maximum likelihood fits to the $p_h$ distribution at each mass point.

No significant excess is observed~\cite{Belle-II:2026tyb}. The largest local excess is $2.95\sigma$ at $m_{X_{\rm{inv}}}=3.28\ \rm{GeV}/c^2$ in the $B^+\to\pi^+X_{\rm{inv}}$ channel, reducing to $0.65\sigma$ global significance. Upper limits at $90\%$ C.L. on ${\cal B}(B\to h X_{\rm{inv}})$ are set in the range $10^{-4}$–$10^{-6}$, depending on channel and mass. These are the most stringent limits to date, and the first results for $B^+\to\pi^+X_{\rm{inv}}$, $B^+\to D_s^+X_{\rm{inv}}$, and $B^0\to\bar{D}^0X_{\rm{inv}}$.

\begin{figure}
    \centering
    \includegraphics[width=0.41\linewidth]{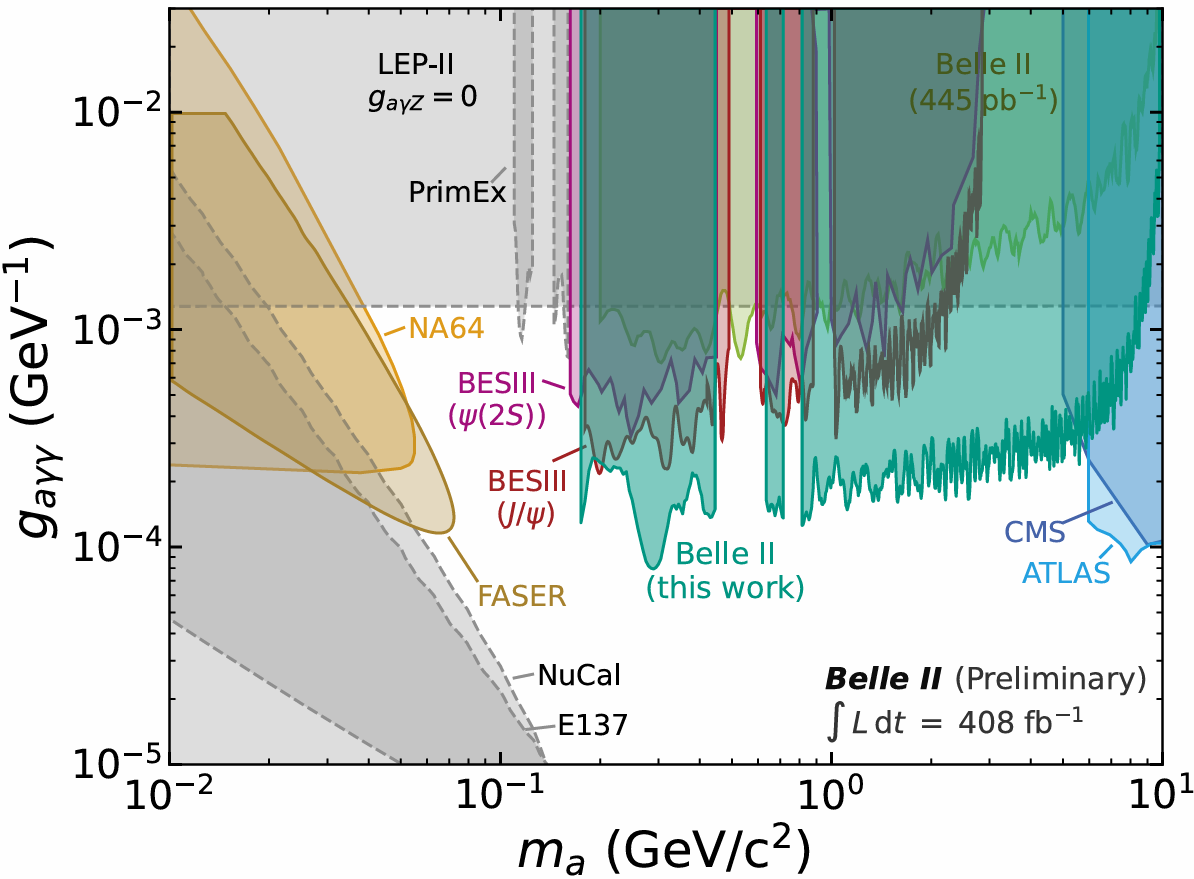}
    \includegraphics[width=0.52\linewidth]{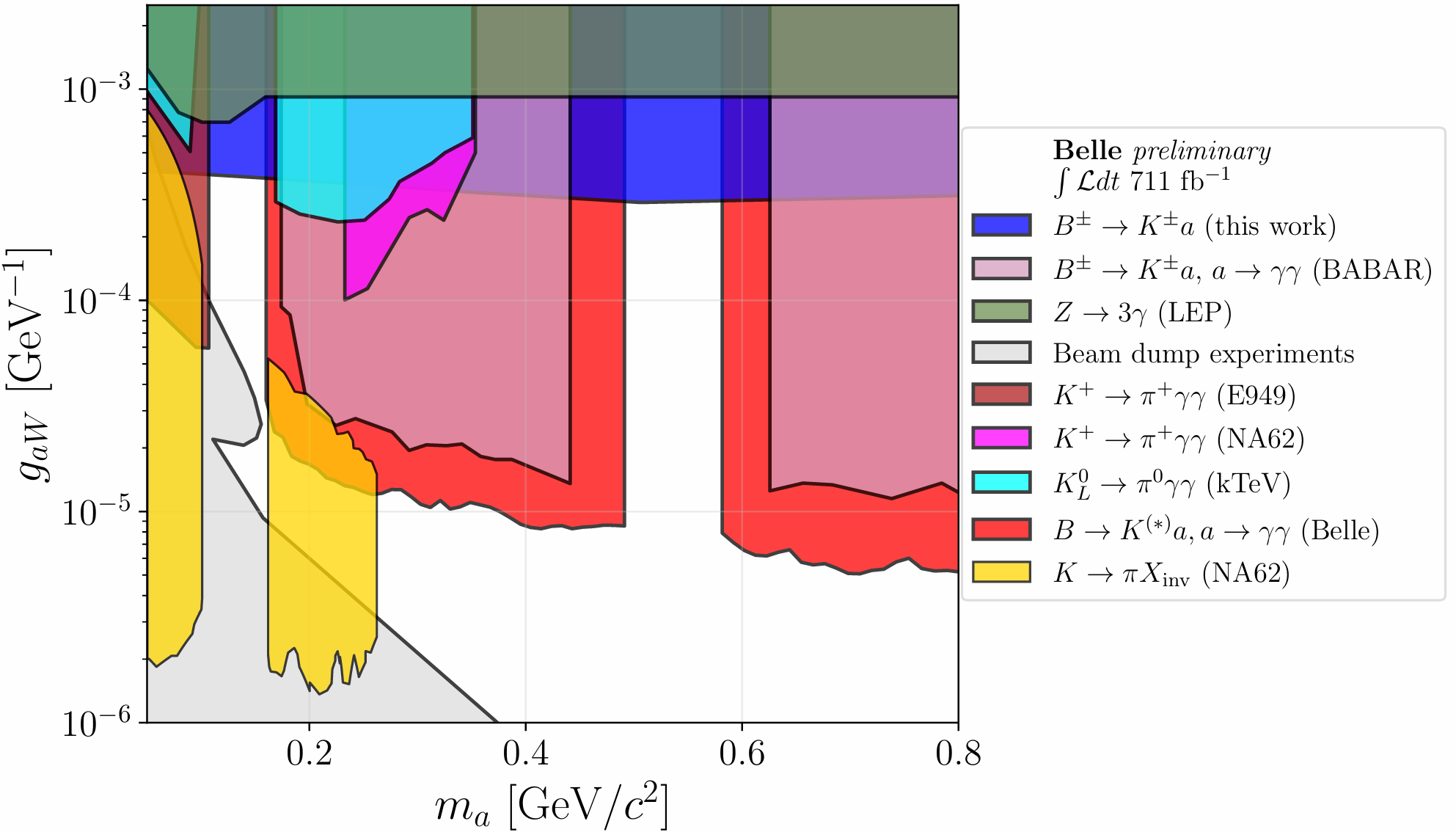}
   \caption{(Left) Exclusion region (95\% C.L.) in the plane of the ALP-photon coupling $g_{a\gamma\gamma}$ and ALP mass $m_a$ from this work (teal) and previous constraints. 
   (Right) Limits on the coupling strength $g_{aW}$ assuming the ALP decays via $a\to\gamma\gamma$.}
    \label{Dark}
\end{figure}

The limits are further interpreted in terms of ALP–$W$ coupling $g_{aW}$, as shown in the right plot of the Fig.~\ref{Dark}, dark scalar–Higgs mixing $\sin\theta$, $B$-mesogenesis~\cite{Elor:2018twp,Alonso-Alvarez:2021qfd}, and RPV SUSY~\cite{Dib:2022ppx}. For example, $B^+\to p X_{\rm{inv}}$ excludes $B$-mesogenesis for dark baryon masses above $2.5\ \rm{GeV}/c^2$ and in the range $2.15$–$2.3\ \rm{GeV}/c^2$.

\section{Summary}
Belle and Belle~II, operating at the $\Upsilon(4S)$ resonance, provide a clean and high-luminosity environment for precision studies of tau leptons and searches for dark sector particles.

In tau physics, Belle~II has performed a first measurement of CP violation in $\tau^- \to \pi^- K_S^0 \nu_\tau$, which results to be consistent with the SM predictions, and  a search for LFV $\tau^- \to \ell^- \eta$, setting world-leading upper limits.

For the dark sector, searches for ALPs in $e^+e^- \to \gamma a$ using Belle~II data and for FIPs in $B \to h X_{\rm inv}$ using Belle data provide the most restrictive constraints to date.

The Belle and Belle II experiments are at the forefront of $\tau$ and dark sector searches, providing world-leading constraints and unique sensitivity. New results, with improved analyses and larger data samples, are expected to further enhance the sensitivity to new physics.


\section*{References}
\bibliography{moriond}

\end{document}